\documentclass{aa}
\usepackage{graphicx}
\begin{document}
   \title{The field brown dwarf LP 944-20 and the Castor moving group}

   \author{I. Ribas}


   \institute{Departament d'Astronomia i Meteorologia, Av. Diagonal, 647,
              E-08028 Barcelona, Spain\\
              \email{iribas@am.ub.es}
}

   \date{Received ; accepted}

   \abstract{A reliable age estimation for the field brown dwarf
\object{LP 944-20} is accomplished through the analysis of its kinematic
properties. The space velocities of this star strongly suggest its
membership in the so-called Castor moving group. LP 944-20 can be sensibly
assumed to have the group's age, which is estimated to be $\sim$320 Myr,
and metal content, which is found to be roughly solar. With these new
constrains and the available photometry and lithium abundance, current
brown dwarf models are put to a test. Using the IR magnitudes and the
lithium diagnostics, the models are able to provide a reasonable
description of the brown dwarf's properties (to within a few sigma) but
yield an age which is roughly 50\% larger than our estimate. Possible
reasons for this discrepancy are discussed.

            }

   \maketitle
%

\section{Introduction}

Brown dwarf (BD) research has made impressive progress after the first
conclusive identification of a BD in 1995 (Nakajima et al. \cite{NOK95}).
These objects, which populate the gap between low-mass stars and planets,
are now observationally quite well characterized, especially with regard
to their radiative and spectroscopic properties (see, e.g., Kirkpatrick et
al. \cite{KRL99}; Basri \cite{B00}). Also, recent theoretical models,
which most notably include dust formation and opacity, have evolved
significantly and are thought to provide a realistic picture of BDs
(Chabrier \& Baraffe \cite{CB00}). However, some of the most important
fundamental quantities, such as masses and ages, have not been directly
determined for any of the {\em bona-fide} field BDs. Obviously, both mass
and age are of crucial importance to assess the ability of theoretical
models to make a trustworthy description of BD's properties. Only by
limiting the number of free parameters in the comparison will we be able
to critically analyze the performance of BD models. Unfortunately, masses
can only be determined in a direct manner (i.e. dynamically) for BDs in
binary systems, which are still very scarce. Age determination for BDs is
even more challenging, and indirect indications from membership in open
clusters or from stellar companions are the most sensible approaches.

Given the situation, one should take advantage of any opportunity to make
reliable estimations of masses and ages of BDs. The field BD LP 944-20, as
we show below, is one of the rare cases that allow for such estimation.
The faint object \object{LP 944-20}, with spectral type M9~V, was first
included in a high-proper motion star catalogue by Luyten \& Kowal
(\cite{LK75}). It was not until much later that Tinney (\cite{T98};
hereafter T98) first established its substellar nature through the
so-called Li test. During its short history as a confirmed BD, LP 944-20
has been observed at various wavelengths and holds the honour of being the
first field substellar object to have been detected both in the radio and
mid-IR domains (Berger et al. \cite{BBB01}; Apai et al. \cite{APH02}).
Also, LP 944-20 has been seen to experience X-ray flares but appears to
have a low X-ray quiescent flux (Rutledge et al. \cite{RBM00})

As a well-observed BD, LP 944-20 can serve as a valuable benchmark to
evaluate the performance of theoretical models. To do so, one needs to
constrain its physical properties as accurately as possible. Apparently LP
944-20, as an isolated object, would present a big challenge for the
determination of its age. In this paper, we employ its kinematic
properties to infer the object's age and metal content and use this
information to carry out a test of current BD models.

\section{The kinematics of LP 944-20 and the Castor moving group} \label{seckin}

The space velocities $(U,V,W)$ of a star are readily computed from its
position, proper motions, radial velocity and distance. In the case of LP
944-20, we have adopted the position, radial velocity and distance from
Tinney (\cite{T96}) and T98. In addition, values for the proper motions
have been provided by Luyten (\cite{L79}) and Tinney (\cite{T96}).
Surprisingly, the difference between these two values is very large,
amounting to over 10$\sigma$. Since the effect on the computed space
velocities is fairly small, we will not discuss possible reasons for the
discrepancy, but the large parallax ($\sim$200 mas) of LP 944-20 or a
typographical error might be responsible. Instead of favoring one of the
values, we have listed the space velocities resulting from the two
different values of the proper motion in Table \ref{tabuvw} (Case I:
Tinney \cite{T96}; Case II: Luyten \cite{L79}). Note that we have followed
the convention where positive values of $U$, $V$, and $W$ indicate
velocities towards the galactic center, galactic rotation and north
galactic pole, respectively.

\begin{table}[!t]
   \caption[]{Astrometric and kinematic data for LP 944-20. Space
velocities in Case I have been computed using the proper motion in Tinney
(\cite{T96}), while Case II makes use of the proper motion in the NLTT
Catalogue of Luyten (\cite{L79}).}
      \label{tabuvw}
      \begin{center}
      \begin{tabular}{lcccc}
         \hline
\multicolumn{1}{c}{Parameter}&&\multicolumn{1}{c}{Case I}&&\multicolumn{1}{c}{Case II}\\
            \hline
$\mu_{\alpha} \cos \delta$ (mas yr$^{-1}$)&&324.0                 &&210.2        \\
$\mu_{\delta}$ (mas yr$^{-1}$)            &&295.9                 &&259.6        \\
$d$ (pc)                                  &&    5.04$\pm$0.11     &&    5.04$\pm$0.11   \\
$v_{\rm r}$ (km s$^{-1}$)                 &&10.0$\pm$2.0          &&10.0$\pm$2.0         \\
$U$ (km s$^{-1}$)                         &&\llap{$-$}12.6$\pm$0.8&&\llap{$-$}10.7$\pm$0.8      \\
$V$ (km s$^{-1}$)                         &&$-$6.1$\pm$1.4        &&$-$4.8$\pm$1.4       \\
$W$ (km s$^{-1}$)                         &&$-$3.5$\pm$1.9        &&$-$5.1$\pm$1.9       \\    
\hline
    \end{tabular}
    \end{center}
\end{table}

The existence of a group of stars kinematically linked to the sextuple
system Castor was first proposed by Anosova \& Orlov (\cite{AO91}). Some
18 stars with spectral types between A1 V and M6 Ve were suggested to be
members of the moving group. Barrado y Navascu\'es (\cite{B98}) later
revisited the situation and provided a new list of Castor moving group
candidates based upon more restrictive criteria such as kinematics,
position in H-R diagram and lithium abundance. Barrado y Navascu\'es lists
a total of 17 possible members of the moving group. More recently, Montes
et al. (\cite{MLG01}) proposed the membership of several additional late
type stars in the moving group using purely kinematic criteria. With these
lists as starting point, we have performed a search in the literature and
used our own observations to compile $VRI$ photometry. The measurements
are presented in Table \ref{tabmgm}. We have further completed the list by
adding the components of the quadruple system \object{GJ 2069}, which also
share the space velocities of the group members. The implications of the
GJ 2069 membership in the moving group are discussed in Ribas
(\cite{R02}).

\begin{table}[!t]
   \caption[]{Photometry for the Castor moving group members in Barrado y
Navascu\'es (\cite{B98}) and Montes et al. (\cite{MLG01}) with additional 
members suggested in this work.}
      \label{tabmgm}
\scriptsize
      \begin{center}
      \begin{tabular}{lllllcl}
\hline
Name                   &SP      &$V$          &$I_{\rm C}$  &\tiny{$\!\!\!\!(R-I)_{\rm C}\!\!\!\!\!\!$}&$d$ (pc)$\!\!\!$&\multicolumn{1}{c}{$M_{\rm I}$}\\
\hline
\object{Gl 20         }&A7 V    & 3.933       & 3.734       &          0.096&   23.5&        1.877       \\
\object{Gl 217.1      }&A2 Vann & 3.542       & 3.436       &          0.049&   21.5&        1.772       \\
\object{Gl 226.2      }&K8 V    & 9.75        & 8.26        &          0.66 &   24.9&        6.276       \\
\object{Gl 255 A      }&F8 IV-V & 6.795       & 6.28        &          0.250&   43.2&        3.10        \\
\object{Gl 255 B      }&--      & 7.209       & 6.67        &          0.265&   43.2&        3.49        \\
\object{Gl 278 A      }&A1 V    & 1.93        & 1.915       &          0.008&   14.9&        1.042       \\
\object{Gl 278 B      }&A5 Vm   & 2.93        & 2.775       &          0.084&   14.9&        1.902       \\
\object{Gl 278 C      }&M1 V    & 9.074       & 7.155       &          0.993&   14.9&        7.035$^{1}$$\!\!\!$ \\
\object{Gl 351 A}$^{2}$&F2 IV   & 3.887       & 3.454       &          0.214&   18.6&        2.111       \\
\object{GJ 521.2 A    }&F7.7 V  & 6.34        & 5.86:       &          0.22:&   25.2&        3.85:       \\
\object{Gl 564.1}$^{2}$&A3 IV   & 2.747       & 2.591       &          0.078&   23.7&        0.720       \\
\object{Gl 721        }&A0 V    & 0.034       & 0.039       &          0.004&   ~7.8&        0.590       \\
\object{Gl 755        }&G5 V    & 6.477       & 5.792       &          0.335&   20.9&        4.195       \\
\object{Gl 826}$^{2}$  &A7 IV-V & 2.459       & 2.195       &          0.120&   15.0&        1.320       \\
\object{GJ 842.2      }&M0.5 V  &\llap{1}0.57 & 8.69        &          0.97 &   20.9&        7.09        \\
\object{Gl 879        }&K4 V    & 6.485       & 5.283       &          0.540&   ~7.6&        5.868       \\
\object{Gl 881        }&A3 V    & 1.155       & 1.077       &          0.025&   ~7.7&        1.647       \\
\object{Gl 896 A      }&M3.5 V  &\llap{1}0.26 & 7.635       &          1.516&   ~6.2&        8.8$^{1}$   \\
\object{Gl 896 B      }&M4.5 V  &\llap{1}2.47 & 9.296       &          1.873&   ~6.2&\llap{1}0.45$^{1}$  \\
\object{GJ 1111       }&M6.5 V  &\llap{1}4.79 &\llap{1}0.53 &          2.26 &   ~3.6&\llap{1}2.73        \\
\object{GJ 2069 Aa    }&M3.5 V  &\llap{1}2.49 & 9.71        &          1.59 &   12.8&        9.17        \\
\object{GJ 2069 Ab    }&M3.5 V  &\llap{1}2.82 & 9.99        &          1.62 &   12.8&        9.45        \\
\object{GJ 2069 Ba    }&M4.5 V  &\llap{1}3.60 &\llap{1}0.68 &          1.66 &   12.8&\llap{1}0.14        \\
\object{GJ 2069 Bb    }&M4.5 V  &\llap{1}4.94 &\llap{1}1.63 &          1.93 &   12.8&\llap{1}1.09        \\
\object{GJ 3633       }&K0 V    & 7.34        & --          &          --   &   17.5&       --           \\
\object{HD 77825      }&K2 V    & 8.78        & 7.82        &          0.48 &   28.1&        5.58        \\
\object{LP 944-20     }&M9 V    &\llap{1}9.2: &\llap{1}4.16 &          2.94 &   ~5.0&\llap{1}5.65        \\
\hline
    \end{tabular}
    \end{center}
\vspace{-2mm}
$^{1}$ Corrected for duplicity.\\
$^{2}$ Possibly not a member. See text.
\end{table}

\begin{figure}
\centering
\includegraphics[width=\columnwidth]{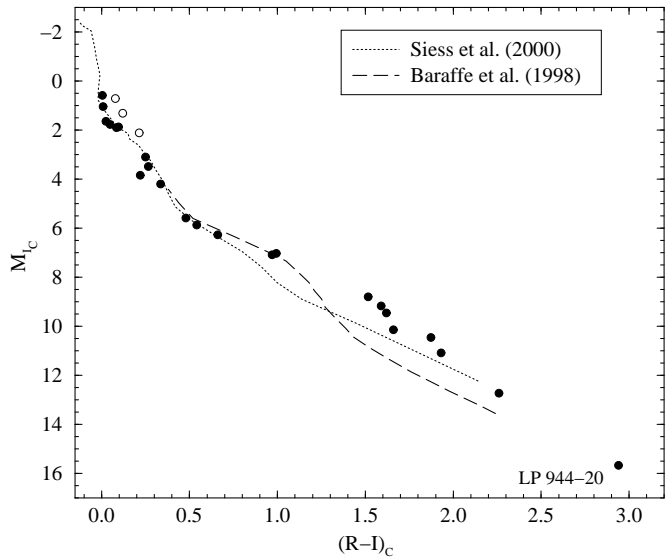}
\caption{Color-magnitude diagram for the Castor moving group members with
$R$ and $I$ photometry. Also plotted are two isochrones with an age of 370
Myr computed using the models of Baraffe et al. (\cite{BCAH98}) and Siess
et al. (\cite{SDF00}). Three early-type stars that deviate significantly
from the isochrone are represented by empty circles and possibly do not
belong in the moving group (see Table \ref{tabmgm} for identification).
The systematic deviations for $(R-I)$ colors greater than 1 (M stars)
are likely to arise from shortcomings of current atmosphere models (see
text for further information).
           \label{figCas}}
\end{figure}

A color magnitude plot of all stars in Table \ref{tabmgm} is shown in
Fig. \ref{figCas}. Also plotted are two isochrones for the estimated age
of the moving group (see next section) computed using the models by Siess
et al. (\cite{SDF00}) and Baraffe et al. (\cite{BCAH98}). Using this
diagram we have further refined the membership list by isolating and
eventually excluding three stars that appear to deviate significantly
above the isochrone (empty circles in Fig. \ref{figCas}). Note that this
criterion can only be applied to stars that are massive enough (i.e. early
spectral types) to experience evolutionary changes in timescales of a few
100~Myr. All the excluded stars have luminosity class IV, which agrees
with our claim that these are older and more evolved than the Castor
moving group members. The agreement of the theoretical isochrones and the
sequence defined by the stars hotter than spectral class M is excellent.
One of the deviating objects with the earliest spectral type is the
double-lined eclipsing binary \object{YY Gem} (\object{Gl 278 C}), which
is composed of a pair of M1 V stars, and was analyzed in detail by Torres
\& Ribas (\cite{TR02}). The authors found that current stellar models fail
to reproduce the observed physical properties of the stars by up to 20\%.
This same situation is also likely to occur for later spectral types.
Thus, the systematic deviation that arises for spectral types M is not
surprising.

With the updated list of members and the individual velocities in Barrado
y Navascu\'es (\cite{B98}) and Montes et al. (\cite{MLG01}) we have
computed mean velocities for the moving group, which are $<U>=-10.6\pm3.7$
km~s$^{-1}$, $<V>=-6.8\pm2.3$ km~s$^{-1}$, and $<W>=-9.4\pm2.1$
km~s$^{-1}$, i.e. ($-10.6$, $-6.8$, $-9.4$) km~s$^{-1}$. The velocities
that we find for LP 944-20 (see Table \ref{tabuvw}) are in good agreement
with the mean values for the Castor moving group. The most discrepant
velocity component is $W$, with a difference of 4--6 km~s$^{-1}$ but still
marginally compatible within the error bars. Both $U$ and $V$ velocities
are well within the uncertainties of the mean values found for the moving
group. Also, we report that LP 944-20 fulfills the two moving group
membership criteria described in Montes et al. (\cite{MLG01}). The total
space velocity difference between LP 944-20 and the Castor moving group
average is 4.7--6.3 km~s$^{-1}$ (depending on the velocities adopted for
LP 944-20; see Table \ref{tabuvw}). This difference is consistent with the
membership of LP 944-20 in the Castor moving group given its relatively
old age (see Sect. \ref{secall}) and the expected peculiar velocities of
the members due to disk heating. A work with similar goals to ours was
carried out by Ruiz et al. (\cite{RTR91}), who studied the membership of
the BD candidate \object{ESO 207-61} in the Hyades moving group. The
difference of space velocities between the BD candidate and the mean
velocity of the moving group was found to be over 9 km~s$^{-1}$, yet these
authors concluded that ESO 207-61 was a likely member of the Hyades moving
group. The case for LP 944-20 is even stronger.

Several studies have been published in recent years (Sabas \cite{S97};  
Dehnen \cite{D98}; Asiain et al. \cite{AFT99}) dealing with the kinematic
analysis of large stellar samples, which have led to the detection of
numerous moving groups. Interestingly, all these studies report stellar
kinematic groups with velocities very similar to those we find for the
Castor moving group: Group ``D'' found by Sabas (\cite{S97}) has velocity
components ($-12.5$,$-6.2$,$-8.3$) km~s$^{-1}$, Dehnen (\cite{D98}) gives
($-10$,$-5$,$-8$) km~s$^{-1}$ for his Group 4, and Asiain et al.  
(\cite{AFT99}) obtain ($-10.5$,$-5.0$,$-7.5$) km~s$^{-1}$ for their group
``C1''.  All these, often referred to as the Coma Berenices group, can
possibly be identified as the Castor moving group with the small velocity
differences arising from the various stellar samples considered and
analysis methods used. Interestingly, the Castor moving group has quite
different velocities to those of all the other young (age$<$600 Myr)
kinematic groups reported by these publications (see also Montes et al.
\cite{MLG01}). Thus, the similarity (within 1$\sigma$) between the
velocities of LP 944-20 and those listed above reduces the possibility of
misclassification strengthens our case for the membership of LP 944-20 in
the Castor moving group.

\section{The age and metallicity of LP 944-20} \label{secall}

LP 944-20 became one of the first {\em bona-fide} BDs when T98 reported
the detection of lithium in the spectrum, thus confirming its substellar
nature. T98 compared the equivalent width of the Li~{\sc i} $\lambda$6708
line with theoretical models and estimated an age between 475 and 650 Myr,
and a mass of 0.056--0.064~M$_{\odot}$. Other independent tests seemed to
roughly confirm an intermediate age for this object.

The membership of LP 944-20 in the Castor moving group discussed in the
previous section opens a new perspective. Indeed, it is sensible to assume
that LP 944-20 has the same age and chemical composition as the rest of
the group members -- similarly to what is routinely done in stellar
cluster studies. Barrado y Navascu\'es (\cite{B98}) used criteria based on
kinematics, isochrone fitting, and lithium abundances to estimate the age
of the moving group to be 200$\pm$100 Myr. Asiain et al.  (\cite{AFT99})  
estimated the age of their ``C1'' group (which we identify as the Castor
moving group) from isochronal fits and obtained 400$\pm$200 Myr. Using a
very similar procedure, Sabas (\cite{S97}) derived an age of 320$\pm$150
Myr for her group ``D'' (also identified as the Castor moving group). The
study by Torres \& Ribas (\cite{TR02}) analyzed in detail the Castor
sextuple system and obtained an age of 370$\pm$50 Myr and a nearly solar
value for the metallicity.  A further independent estimate is possible
through the study of the activity level of the solar analog \object{Gl
755} (spectral type G5 V). Stars narrowly confined in spectral type
between G0 V and G5 V have well studied correlations between their X-ray
luminosities and ages (see Dorren et al. \cite{DGG95}). Using the X-ray
luminosity for Gl 755 in H\"unsch et al. (\cite{HSSV99})  and the
relationship of G\"udel et al. (\cite{GGS97}) we derive a lower limit for
its age of 200 Myr. Note that only a lower limit can be obtained because
Gl 755 is a poorly-understood close binary and a fraction of the X-ray
emission could be coming from an unseen companion (Gaidos \cite{G98}).

With these considerations in mind, we have adopted an age of 320$\pm$80
Myr and a metal content of $Z=Z_{\odot}$ for the moving group members. Our
value for the age is therefore lower than the model-dependent estimate by
T98. Even though sometimes overlooked, chemical composition is also an
important parameter that has a critical influence on the evolution of
stars and BDs. Not only we have just obtained an independent estimate of
the age of LP 944-20 but also we have information on its metallicity.
Under these circumstances a test of current BD models is possible. Also,
the estimated age of LP 944-20 is nicely placed between the Pleiades and
Hyades and allows for an analysis of models in an age domain so far
unexplored.

\begin{table}[!t]
   \caption[]{Red and IR photometry for LP 944-20. The transformation from
the 2MASS to the CIT systems has been carried out by using expressions in
Carpenter (\cite{C01}).}
      \label{tabphot}
      \begin{center}
      \begin{tabular}{lrlr}
         \hline
\multicolumn{4}{c}{Red and IR photometry (mag)}\\
            \hline
$J_{\rm 2MASS}=          $&$10.75\pm0.03$&$I_{\rm C}=              $&$14.16\pm0.04$\\
$H_{\rm 2MASS}=          $&$10.02\pm0.03$&$(m-M)=                  $&$-1.49\pm0.05$\\
${K_{\rm S}}_{\rm 2MASS}=$&$~9.52\pm0.03$&${M_{\rm I}}_{\rm C}=    $&$15.65\pm0.04$\\
$J_{\rm CIT}=            $&$10.72\pm0.03$&${M_{\rm K}}_{\rm CIT}=  $&$11.04\pm0.06$\\
$H_{\rm CIT}=            $&$10.00\pm0.03$&\scriptsize{$(I_{\rm C}-K_{\rm CIT})=$}&$~4.61\pm0.05$\\
$K_{\rm CIT}=            $&$~9.55\pm0.03$&&\\
\hline
    \end{tabular}
    \end{center}
\end{table}

\section{Comparison with brown dwarf models} \label{secmod}

Several theoretical models for substellar masses are currently available
in the literature. These include the models of D'Antona \& Mazzitelli
(\cite{DM94}), Burrows et al. (\cite{Bea97}), Baraffe et al.
(\cite{BCAH98}), and Chabrier et al. (\cite{CBAH00}). With both the age
and the metal content known, the mass and temperature of LP 944-20 can be
readily estimated by interpolating these models at, for instance, the
observed bolometric luminosity (to avoid employing uncertain color
transformations for the time being). The luminosity of LP 944-20 follows
from its absolute $K$ magnitude (in Table \ref{tabphot}) and a bolometric
correction (${BC_{\rm K}}_{\rm 2MASS}=+3.21\pm0.07$ mag) computed using
the atmosphere models by Allard et al. (\cite{AHA01}). We obtained a value
of $\log(L/{\rm L}_{\odot})=-3.79\pm0.04$. Interpolation in the model
grids yielded values for the mass and effective temperature presented in
Table \ref{tabmteff}. Mass estimates cluster around 0.05~M$_{\odot}$ with
a scatter of about 6\% and the effective temperature comes out to be
nearly 2000~K with a scatter of 2.5\%. Note that this value is
significantly smaller than the effective temperature estimated by Basri et
al. (\cite{Bea00}) for LP 944-20, which was found to be $T_{\rm
eff}\approx2400$~K from spectral line fits. Also, Luhman (\cite{L99})
assigns an effective temperature of 2400~K to dwarf M9 objects by
extrapolating the scale of Leggett et al. (\cite{Lea96}). These results
suggest the existence of significant disagreements between the effective
temperatures predicted by the models and the empirical scales. Our value
for the mass is somewhat lower than the estimate by T98 who obtained
0.056--0.064~M$_{\odot}$ using similar BD models. The reason for this
higher mass can be found in the larger value for the age adopted by T98.

\begin{table}[!t]
   \caption[]{Mass and effective temperature derived from the observed age
and luminosity of LP 944-20 for several BD models. The
interpolation parameter was the bolometric luminosity.}
      \label{tabmteff}
      \begin{center}
      \begin{tabular}{lll}
         \hline
\multicolumn{1}{c}{Models}   &\multicolumn{1}{c}{Mass (M$_{\odot}$)} & \multicolumn{1}{c}{T$_{\rm eff}$ (K)} \\
            \hline
D'Antona \& Mazzitelli (\cite{DM94})&    0.055             &   2060              \\
Burrows et al. (\cite{Bea97})       &    0.050--0.055      &   2000--2100        \\
Baraffe et al. (\cite{BCAH98})      &    0.049             &   2000              \\
Chabrier et al. (\cite{CBAH00})     &    0.049             &   1970              \\
\hline
    \end{tabular}
    \end{center}
\end{table}

After this rough comparison with model data, we proceeded one step further
and performed a more critical test of models using the available
photometric information. The recent BD models of Chabrier et al.
(\cite{CBAH00}) were employed for a detailed comparison because these
provide a complete set of optical/IR magnitudes, lithium abundances and
bolometric luminosities for a range of masses and ages. Also these models
incorporate modern input physics and atmospheres that include dust
formation.

The available data for LP 944-20 to perform a comparison with models are
accurate IR magnitudes and colors, and a measurement of the Li~{\sc i}
$\lambda$6708 equivalent width. IR photometry of LP 944-20 is available
from the 2MASS Point Source Catalogue Second Incremental Release
(Skrutskie et al. \cite{Sea97}) and is presented in Table \ref{tabphot}.
Also in the table are the IR magnitudes transformed to the CIT system
using expressions in Carpenter (\cite{C01}). This was done because
photometry of BD models in Chabrier et al. (\cite{CBAH00}) is given in the
CIT system.

\begin{figure}
\centering
\includegraphics[width=\columnwidth]{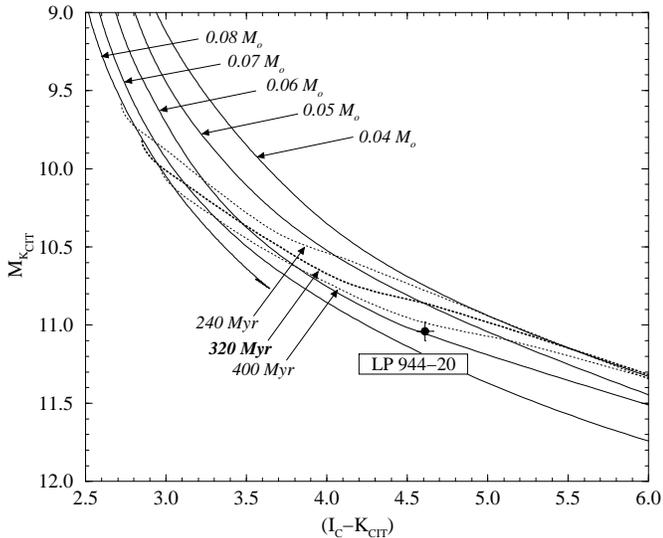}
\caption{Color--magnitude diagram at the BD regime. The
observational data point for LP 944-20 is shown. Evolution tracks (solid
lines) and isochrones (dotted lines) were interpolated from the model grid
of Chabrier et al. (\cite{CBAH00}).}
\label{figTra}
\end{figure}

Figure \ref{figTra} presents a color magnitude diagram at the bottom of
the main sequence. The data point with error bars corresponds to the
photometry of LP 944-20 with its uncertainty. Evolution tracks for several
masses computed by interpolating in the model grids of Chabrier et al.
(\cite{CBAH00}) are represented with solid lines. Isochrones for the
estimated age of LP 944-20 and $\pm$1$\sigma$ were also interpolated from
the models and are shown on the plot as dotted lines. As can be seen, the
observed photometry for LP 944-20 is just marginally compatible with the
model predictions at the upper limit for the age (400 Myr) and some
2.5$\sigma$ below the predicted $K$ magnitude at the adopted age
(320~Myr). Using the photometry alone, BD models would overestimate the
age by predicting a value of about 500~Myr (and a mass of
0.06~M$_{\odot}$). This result indicates that, while models provide a
reasonable general description of the BD's radiative properties, a closer
look indicates that a photometry-based determination of the age for LP
944-20 may be overestimated by as much as 50\%.

A complementary test of models follows from the comparison of the
predicted and observed lithium abundances. As it is well known, the
lithium test has become one of the primary diagnostics to identify an
object as a BD (see Basri \cite{B00} for a review). The test is especially
useful for candidates of spectral type later than M7 because lithium would
be completely depleted should they not be BDs. This is the case of LP
944-20, with a spectral type of M9, where the detection of the Li~{\sc i}
$\lambda$6708 line was reported by T98. However, lithium is also
eventually destroyed in the more massive BDs when their central
temperatures become sufficiently high. The sole detection of lithium in LP
944-20 (and with its known luminosity) proves that its mass is substellar
and provides (model-dependent) upper limits to both the age ($\leq$650
Myr) and the mass ($\leq$0.065~M$_{\odot}$) of the object.

More quantitatively, the Li~{\sc i} $\lambda$6708 equivalent width
measured by T98 seems to indicate that lithium has already been partially
depleted in LP 944-20, with a current abundance ratio of $\log [n({\rm
Li})/n_{\circ}({\rm Li})]=-2.5\pm1.0$. Using our estimate for the age
(320$\pm$80 Myr) and interpolating in the model grids of Chabrier et al.
(\cite{CBAH00}) leads to a mass of 0.063$\pm$0.001~M$_{\odot}$ and a
bolometric luminosity of $\log(L/{\rm L}_{\odot})=-3.50\pm0.14$. Note that
this latter value is inconsistent with the observed luminosity of
$\log(L/{\rm L}_{\odot})=-3.79\pm0.04$. Turning the argument around, when
both the observed lithium abundance and luminosity are used to compare
with model predictions one obtains an age of 490$\pm$100 Myr and a mass of
0.061$\pm$0.005~M$_{\odot}$. This was the exact same procedure used by
T98, as illustrated in his figure 2. Again, the results obtained from the
lithium abundance seem to overestimate the age of LP 944-20 by about 50\%.

The apparent inconsistency found when comparing the known age of LP 944-20
with estimates from photometry is not surprising. While atmospheres of
very low mass objects have come a long way, there is still room for
improvement and it is likely that broad band magnitudes and colors can be
refined. In this sense, our result for LP 944-20 may prove useful as it
can define an anchor point for intermediate age BDs The differences
detected when using the observed lithium abundances are somewhat more
disturbing. The lithium depletion mechanism is simple and apparently well
understood. At least four scenarios can be put forward to explain the
observed discrepancy:  {\em 1)} An underestimation of the measured lithium
abundances. LP 944-20 is a relatively fast rotator ($v\sin i=30\pm2.5$ km
s$^{-1}$; Basri et al. \cite{Bea00}) and the spectrum acquired by T98 has
fairly low S/N, which could have resulted in an underestimated equivalent
width. {\em 2)} Current BD models underestimate the lithium destruction
rate. Rotation, which is not currently accounted for by evolution models,
could play an important role (Mart\'{\i}n \& Claret \cite{MC96}). A
tantalizing hint that favors this scenario is the nearly constant 50\% age
differential found between the ``classical'' ages of young clusters and
the lithium-based determinations (see Basri \cite{B00} for a summary).
This is the exact same differential we find for LP 944-20. {\em 3)} Dobbie
et al. (\cite{DPJ02}) have recently suggested that theoretical models
might systematically underestimate BD masses. Both the temperature and Li
abundance discrepancies described above could be diminished or resolved if
BD masses predicted by models were larger. {\em 4)} Obviously, a
possibility still remains that LP 944-20 is not a member of the Castor
moving group and that its age and mass are indeed around 500~Myr and
0.06~M$_{\odot}$, respectively. This cannot be ruled out completely
because the velocities of the Castor moving group are quite similar to
those of the Local Standard of Rest (Dehnen \& Binney \cite{DB98}).

\section{Conclusions}

In this paper we exploit the probable membership of the lithium BD LP
944-20 in the Castor moving group. Two otherwise unknown parameters, the
age and metal content, can be obtained for the moving group from a variety
of methods (isochrones, activity-age relationship, lithium abundance
measurements, etc) and then adopted for LP 944-20. Very importantly, the
knowledge of the age and metallicity reduces the number of free parameters
and a critical test of BD models becomes possible. Our results show that
models are able to reproduce the observed properties of LP 944-20 to
within a few sigma, but some discrepancies were detected. In particular,
the analysis of the IR color-magnitude diagram revealed that, with current
atmosphere models, the age of LP 944-20 obtained from this diagnostic
would be overestimated by as much as 50\%. A similar result was found when
considering the age predicted by the models when using the measured
lithium abundance.

Our age estimate for LP 944-20 not only serves as a constraint to evaluate
theoretical models but it also bears on the question of the formation
mechanism of BDs. Apai et al. (\cite{APH02}) recently reported the non
detection of mid-IR excess in LP 944-20 thus indicating the lack of warm
dust around it. The new age we have determined sets a tight constraint on
the disk dissipation time that BD formation models will have to address.

The Castor moving group could be very important because most of its
members lie within 5--20 pc of the Sun. As a result, the main sequence of
the group can be relatively easily followed down to very low masses. As an
intermediate-age association (320~Myr), the Castor moving group can prove
very useful to study BDs in an evolutionary stage beyond that of the
Pleiades objects. With this goal in mind, we are currently investigating
the possible membership of other nearby BD candidates in this moving
group. A very promising object is the M8.5 dwarf \object{2MASSI
J0335020+234235} for which Reid et al. (\cite{RKL02}) determined space
velocities of $U=-13.0\pm1.3$ km~s$^{-1}$, $V=-4.8\pm0.4$ km~s$^{-1}$, and
$W=-4.7\pm0.1$ km~s$^{-1}$. These velocities are very similar to those of
LP 944-20 and the Castor moving group average (see Table \ref{tabuvw} and
Sect. \ref{seckin}). Interestingly, Reid et al. reported the detection of
lithium in 2MASSI J0335020+234235 thus proving its BD status. According to
our results, this BD could also have an age of about 320 Myr. The analysis
of this and further Castor moving group BDs should be able to shed some
light on the age discrepancy found for LP 944-20. Furthermore, we are also
extending the analysis to other young, nearby moving groups that will
permit the generalization of the study to BDs with a variety of ages and
chemical compositions.

\begin{acknowledgements} 
I am indebted to R. Rebolo, who suggested the idea of attempting a
kinematic age determination for LP 944-20. I gratefully acknowledge F. 
Figueras, D. Fern\'andez and C. Jordi for fruitful discussions, and the
referee, R. F. Jameson, for a number of constructive and useful comments.
This publication makes use of data products from the Two Micron All Sky
Survey, which is a joint project of the University of Massachusetts and
the Infrared Processing and Analysis Center/California Institute of
Technology, funded by the National Aeronautics and Space Administration
and the National Science Foundation. This research has made use of the
SIMBAD database, operated at CDS, Strasbourg, France. This research has
made use of NASA's Astrophysics Data System.
\end{acknowledgements}


\begin{thebibliography}{}
\bibitem[2001]{AHA01}
Allard, F., Hauschildt, P. H., Alexander, D. R., Tamanai, A., \& Schweitzer,
A. 2001, ApJ, 556, 357
\bibitem[1991]{AO91}
Anosova, J. P., \& Orlov, V. V. 1991, A\&A, 252, 123
\bibitem[1999]{AFT99}
Asiain, R., Figueras, F., Torra, J., \& Chen, B. 1999, A\&A, 341, 427
\bibitem[2002]{APH02}
Apai, D., Pascucci, I., Henning, Th., et al. 2002, ApJ, 573, L115
\bibitem[1998]{BCAH98}
Baraffe, I., Chabrier, G., Allard, F., \& Hauschildt, P. H. 1998, A\&A, 337, 
403
\bibitem[1998]{B98}
Barrado y Navascu\'es, D. 1998, A\&A, 339, 831
\bibitem[2000]{B00}
Basri, G. 2000, ARA\&A, 38, 485
\bibitem[2000]{Bea00}
Basri, G., Mohanty, S., Allard, F., et al. 2000, ApJ, 538, 363
\bibitem[2001]{BBB01}
Berger, E., Ball, S., Becker, K. M., et al. 2001, Nature, 410, 338
\bibitem[1997]{Bea97}
Burrows, A., Marley, M., Hubbard, W. B., et al. 1997, ApJ, 491, 856
\bibitem[2001]{C01}
Carpenter, J. 2001, AJ, 121, 2851
\bibitem[2000]{CB00}
Chabrier, G., \& Baraffe, I. 2000, ARA\&A, 38, 337
\bibitem[2000]{CBAH00}
Chabrier, G., Baraffe, I., Allard, F., \& Hauschildt, P. H. 2000, ApJ, 542, 464
\bibitem[1994]{DM94}
D'Antona, F., \& Mazzitelli, I. 1994, ApJS, 90, 467
\bibitem[1998]{D98}
Dehnen, W. 1998, AJ, 115, 2384
\bibitem[1998]{DB98}
Dehnen, W., \& Binney, J. J. 1998, MNRAS, 298, 387
\bibitem[2002]{DPJ02}
Dobbie, P. D., Pinfield, D. J., Jameson, R. F., \& Hodgkin, S. T. 2002, 
MNRAS, 335, L79
\bibitem[1995]{DGG95}
Dorren, J. D., G\"udel, M., \& Guinan, E. F. 1995, ApJ, 448, 431
\bibitem[1998]{G98}
Gaidos, E. J. 1998, PASP, 110, 1259
\bibitem[1997]{GGS97}
G\"udel, M., Guinan, E. F., \& Skinner, S. L. 1997, ApJ, 483, 947
\bibitem[1999]{HSSV99}
H\"unsch, M., Schmitt, J. H. M. M., Sterzik, M. F., \& Voges, W. 1999, A\&AS, 
135, 319
\bibitem[1999]{KRL99}
Kirkpatrick, J. D., Reid, I. N., Liebert, J., et al. 1999, ApJ, 519, 802
\bibitem[1996]{Lea96}
Leggett, S. K., Allard, F., Berriman, G., Dahn, C. C., \& Hauschildt, P. H. 
1996, ApJS, 104, 117
\bibitem[1999]{L99}
Luhman, K. L. 1999, ApJ, 525, 466
\bibitem[1979]{L79}
Luyten, W. J. 1979, LHS catalogue. A catalogue of stars with proper motions
exceeding 0\farcs5 annually (University of Minnesota, Minneapolis)
\bibitem[1975]{LK75}
Luyten, W. J., \& Kowal, C. T. 1975, Proper Motion Survey with the 48 inch
Schmidt Telescope. XLIII. One Hundred and Six Faint Stars with Large Proper
Motion (University of Minnesota, Minneapolis)
\bibitem[1996]{MC96}
Mart\'{\i}n, E., \& Claret, A. 1996, A\&A, 306, 408
\bibitem[2001]{MLG01}
Montes, D., L\'opez-Santiago, J., G\'alvez, M. C., et al. 2001, MNRAS, 328, 45
\bibitem[1995]{NOK95}
Nakajima, T., Oppenheimer, B. R., Kulkarni, S. R., et al. 1995, Nature, 378, 
463
\bibitem[2002]{RKL02}
Reid, I. N., Kirkpatrick, J. D., Liebert, J., et al. 2002, AJ, 124, 519
\bibitem[2002]{R02}
Ribas, I. 2002, A\&A, (in press) (astro-ph/0211086)
\bibitem[1991]{RTR91}
Ruiz, M. T., Takamiya, M. Y., \& Roth, M. 1991, ApJ, 367, L59
\bibitem[2000]{RBM00}
Rutledge, R. E., Basri, G., Mart\'{\i}n, E. L., \& Bildsten, L. 2000, ApJ,
538, L141
\bibitem[1997]{S97}
Sabas, V. 1997, PhD Thesis, Obs. de Paris
\bibitem[2000]{SDF00}
Siess, L., Dufour, E., \& Forestini, M. 2000, A\&A, 358, 593
\bibitem[1997]{Sea97}
Skrutskie, M. F., Schneider, S. E., Stiening R., et al. 1997, in The Impact
of Large Scale Near-IR Sky Surveys, ed. F. Garz\'on et al. (Kluwer, Dordrecht), 
25
\bibitem[1996]{T96}
Tinney, C. G. 1996, MNRAS, 281, 644
\bibitem[1998]{T98}
Tinney, C. G. 1998, MNRAS, 296, L42 (T98)
\bibitem[2002]{TR02} 
Torres, G.,\& Ribas, I. 2002, ApJ, 567, 1140
\end{thebibliography}
\end{document}